\documentclass[preprint,authoryear,12pt]{elsarticle}

\usepackage{amssymb}

\usepackage[%
a4paper=true,%
breaklinks=true,%
colorlinks=true,%
pdfauthor={Moreno C\'ardenas et al.},%
pdftitle={Template for manuscripts in Advances in Space Research}%
]{hyperref}

\journal{Advances in Space Research}

\def \grado {$^\circ$}

\begin{document}

\begin{frontmatter}

\title{The Grand Aurorae Borealis Seen in Colombia in 1859}

\author[label1]{Freddy Moreno C\'ardenas}
\ead{ceaf@campestre.edu.co}

\author[label1]{Sergio Cristancho S\'anchez}
\author[label2]{Santiago Vargas Dom\'inguez}
\ead{svargasd@unal.edu.co}

\address[label1]{Centro de Estudios Astrof\'isicos, Gimnasio Campestre, Bogot\'a, Colombia}
\address[label2]{Universidad Nacional de Colombia - Sede Bogot\'a - Facultad de Ciencias - Observatorio Astron\'omico - Carrera 45 \# 26-85, Bogot\'a - Colombia}

\begin{abstract}

On Thursday, September 1, 1859, the British astronomer Richard Carrington, for the first time ever, observes a spectacular gleam of visible light on the surface of the solar disk, the photosphere. The Carrington Event, as it is nowadays known by scientists, occurred because of the high solar activity that had visible consequences on Earth, in particular reports of outstanding aurorae activity that amazed thousands of people in the western hemisphere during the dawn of September 2. The geomagnetic storm, generated by the solar-terrestrial event, had such a magnitude that the auroral oval expanded towards the equator, allowing low latitudes, like Panama's 9\grado N, to catch a sight of the aurorae. An expedition was carried out to review several historical reports and books from the northern cities of Colombia allowed the identification of a narrative from Monter\'ia, Colombia (8$^\circ$ 45' N), that describes phenomena resembling those of an aurorae borealis, such as fire-like lights, blazing and dazzling glares, and the appearance of an immense S-like shape in the sky. The very low latitude of the geomagnetic north pole in 1859, the lowest value in over half a millennia, is proposed to have allowed the observations of auroral events at locations closer to the equator, and supports the historical description found in Colombia. The finding of such chronicle represents one of the most complete descriptions of low-latitude sightings of aurorae caused by the Carrington Event. \\

\end{abstract}


\end{frontmatter}

\parindent=0.5 cm

\section{Introduction}

\subsection{The Carrington Event}

\begin{figure*}
  \begin{center}
    \leavevmode
     \includegraphics[width=0.7\linewidth,angle=0,trim = 0mm 0mm 0mm 0mm, clip]{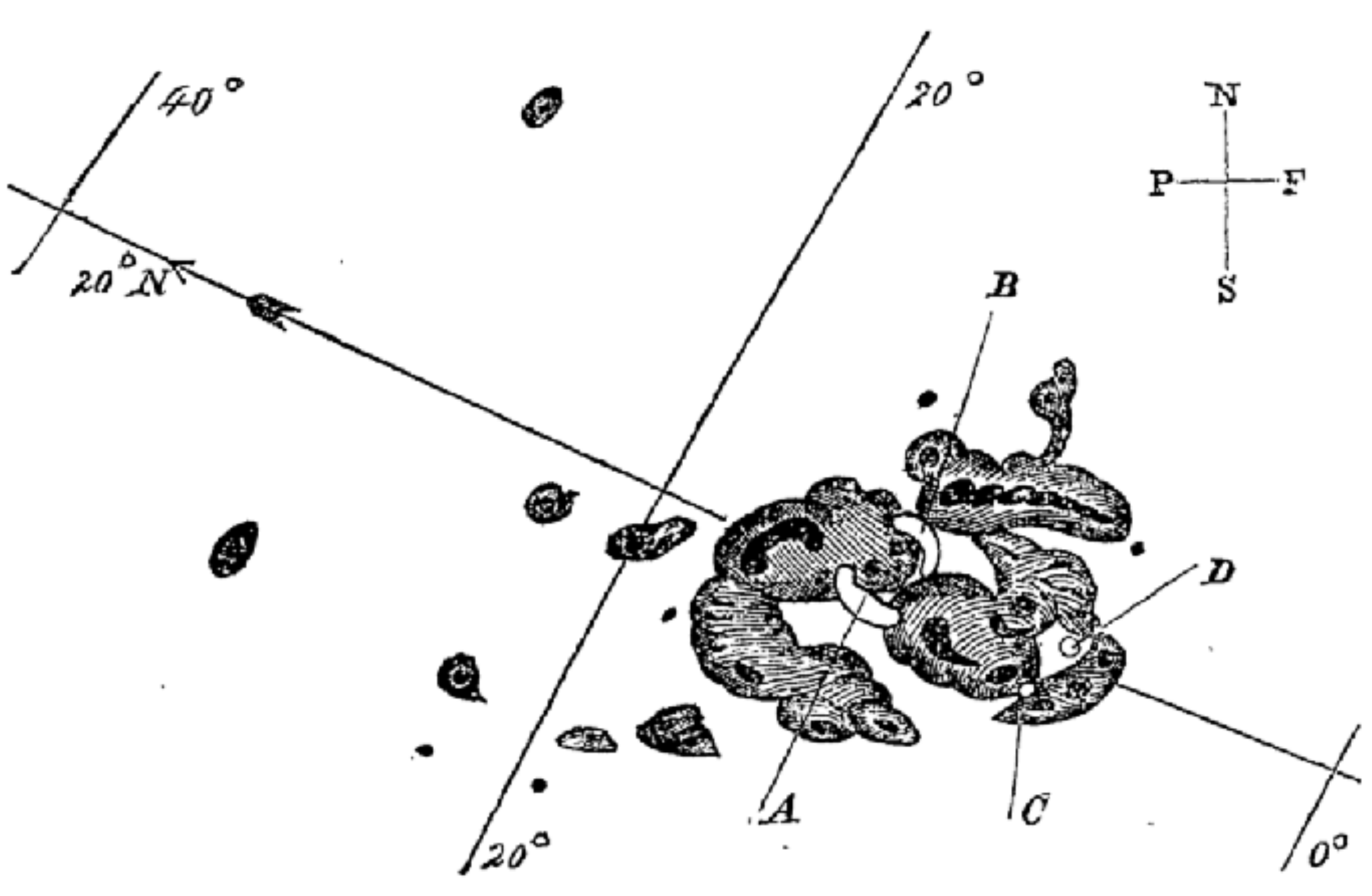}
       \caption{Famous sketch of the solar Active Region 520 and the solar flare (A and B) of September 1, 1859, as drawn by \cite{Carrington1859}.}
     \label{figure1}
  \end{center}
\end{figure*}

On Thursday, September 1, 1859, the British astronomer Richard Carrington described the happenstance, while observing the sunspot group number 520 (Fig.~\ref{figure1})\footnote{Due to the fact that this work is widely based on historical records, we consider important to quote verbatim (\emph{in italics}) part of the sources referenced in the text.}: \emph{The observation of this very splendid group on September 1st has had some notoriety. Mr. Hogdson at Highgate and I at Redhill witnessed and described a singular outbreak of light which lasted about 5 minutes, and moved sensibly over the contour of the spot, an account of which has been sufficiently published by me in the Monthly Notices of the R. A. Society for November, 1859, and since reprinted in the Philos. Trans, Vol. 151, Part III, by Mr. Stewart, in his Memoir on the Great Magnetic Disturbances which extended from August 28th to Sept. 7th.} \citep{Carrington1859, Carrington}.

\begin{figure*}
  \begin{center}
    \leavevmode
      \includegraphics[width=0.9\linewidth,angle=0,trim = 0mm 0mm 0mm 0mm, clip]{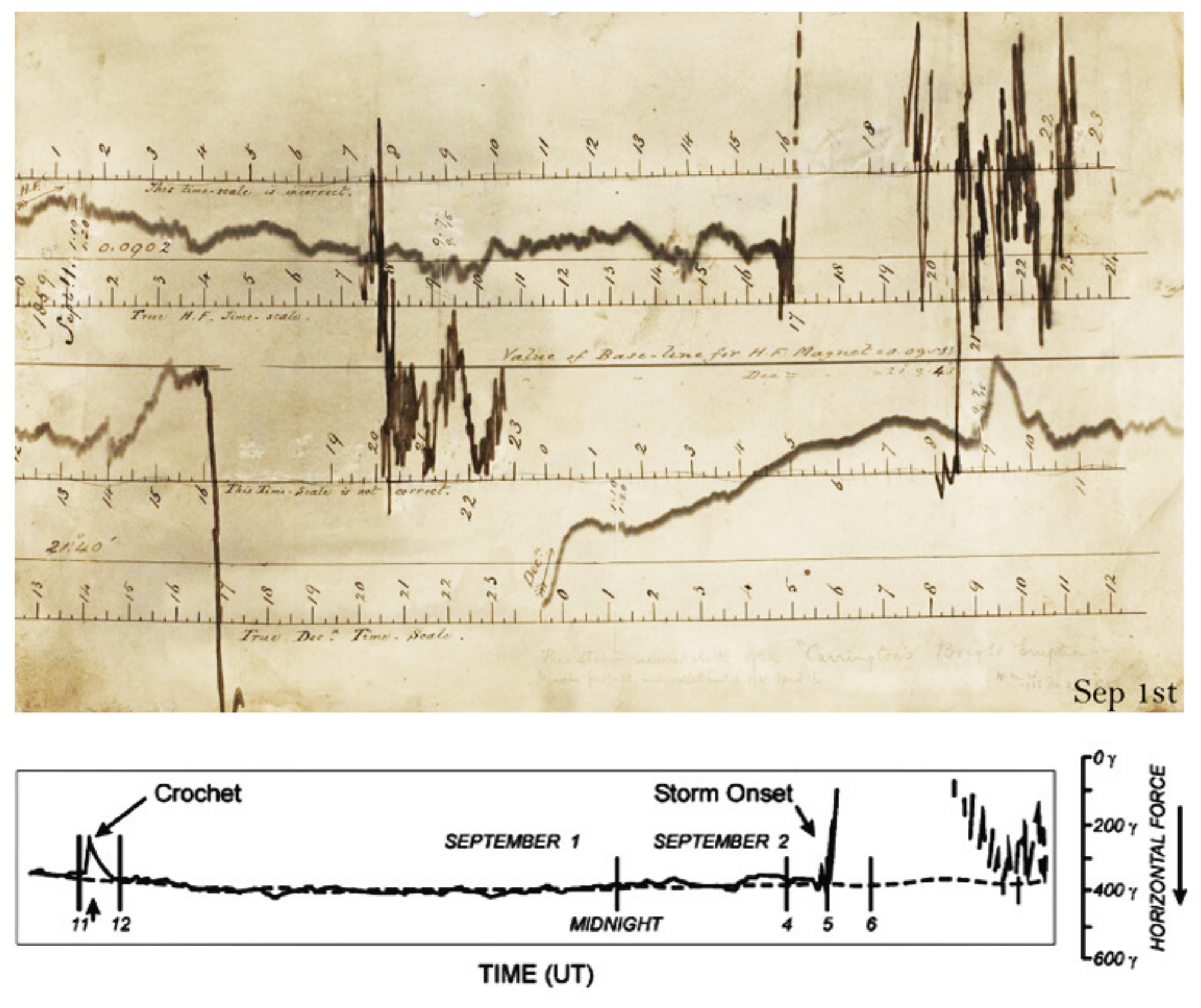}
       \caption{The register of the geomagnetic storm of September 1st, 1859, from the Kew Observatory. \emph{Upper:} Taken from \url{http://www.geomag.bgs.ac.uk/education/carrington.html} and \emph{Lower}: As sketched by \cite{Cliver}, with the indication of some important stages. The intensity of the magnetic field is commonly measured in Gauss (G), or gamma($\gamma$) which is equivalent to a nanotesla (nT, with 1 G = 100.000 nT). See text for details.}
     \label{figure2}
  \end{center}
\end{figure*}

At 11:18 UT, Carrington observed a blast of white light that seemed to be emitted by two points in a group of sunspots; the phenomenon augmented its intensity and adopted a kidney-like form (see Fig.~\ref{figure1}).  Carrington had witnessed the first solar flare ever seen and reported by a human being. The same phenomenon was detected by magnetometers at the Kew Observatory in London (see upper panel in Figure~\ref{figure2}), registering the most important geomagnetic storm announced to date. From August 28 to September 7, the registered data showed a great disturbance, in particular on August 28 and September 2 the most remarkable events  took place, which coincided with the sight of two grand aurorae. Also, another disturbance was registered, at the same hour Carrington and Hogdson saw the flare, in which the horizontal and vertical components of the terrestrial field remained depressed under the normal values for over seven hours \citep{Stewart}.

The first part of the geomagnetic record of Carrington is the mark of the X-flare in the ionosphere, called 'crochet effect' (positive peak). The magnetic recording made at Kew Observatory on September 1 in Fig.~\ref{figure2} shows the turbulence (crochet) at 11:15 UT that matches the same time observing the solar flare seen by Carrington. After that there is the sudden commencement, main phase of the geomagnetic storm, and the recovery phase\footnote{Starting at 8:00 UT, the aurora was seen in Monter\'ia (Colombia).}, as reported in   \cite{Cliver} (see also the lower panel in Fig.~\ref{figure2} of the sketch presented by these authors).  A magnetic crochet, or solar flare effect (SFE) in modern terms, is a type of sudden ionospheric disturbance (SID) caused by flare-induced enhancements of ionospheric E-region currents \citep{Cliver2,Villante}. The geomagnetic storm of 1859 was the first event registered by our civilization on a truly global perspective, and that would not happen again until 1883 when the eruption of the Krakatoa turned dusk crimson red. 

Nonetheless, the phenomenon that was most impressive in the western hemisphere was the auroral activity in the morning of August 29 and September 2 and 3. Newspapers all around the globe informed the unusual aurorae \citep{Green2006}: 
\emph{The light appeared in streams, sometimes of a pure milky whiteness and sometimes of a light crimson ... The crown above, indeed, seemed like a throne of silver, purple and crimson, hung and spread out with curtains or wings of dazzling beauty.} [Washington Daily National Intelligencer, Wednesday, August 31, 1859]\footnote{Text in square brackets denote primary sources from newspapers.}.\\

In Cleveland, Ohio, located in latitude 41\grado N, the following was sighted: \emph{Objects at a distance could be more readily and clearly distinguished than when the moon is at its full ... The whole sky appeared mottled-red, the arrows of fire shooting up from the north, like a terrible bombardment ...} Correspondence of the Journal of Commerce, Cleveland, Ohio. [Washington Daily National Intelligencer, Friday, September 2, 1859]. \\

The aurorae were also observed in the southern hemisphere. The following report was written in Moreton Bay, now Brisbane (Latitude 27.5\grado S) Australia: 
\emph{Most of our readers saw last week, for three nights , commencing after sunset  and lighting up the heavens with a gorgeous hue of red, the Southern Aurora $\cdots$} [Moreton Bay Courier, Wednesday, September 7, 1859].
  
But, not only the environment was disturbed; the telegraphic lines of several countries were affected, leaving them inoperative for several hours: 
\emph{The French telegraph communications at Paris were greatly affected, and on interrupting the circuit of conducting wire strong sparks were observed $\cdots$} [The Illustrated London News, September 24, 1859].
  
In some cases, the consequences of the storm allowed the communication between locations far away, like Boston and Portland, which was the first registration ever of a message transmitted due to the energy supplied by energetic phenomenon \citep{Green2005a}: \emph{The wire was then worked for about two hours without the usual batteries, on the auroral current, working better than with the batteries connected. The current varied, increasing and decreasing alternately, but by graduating the adjustment to the current, a sufficiently steady effect was obtained to work the line very well $\cdots$} [The Daily Chronicle and Sentinel, Augusta, Georgia, Thursday, September 8, 1859].
  
During the most intense geomagnetic storms, when aurorae expand towards the geomagnetic equator, occasionally, these luminous phenomena become visible in tropical latitudes.
\emph{All our exchanges, from the northern coast of the Island of Cuba (from the  southern side we have none so late,) come to us with  glowing descriptions of the recent Aurora Borealis, which appears to have been as bright in the tropics as in the northern zones, and far more interesting $\cdots$} \citep{Green2006}. 
  
The 1859's geomagnetic storm was of such magnitude that its corresponding aurorae was observed in latitudes even lower than Cuba's, like in Jamaica, at 18$^\circ$ N \citep{Nevanlinna}.

\subsection{Historical report from Colombia}

\begin{figure*}
  \begin{center}
    \leavevmode
      \includegraphics[width=0.8\linewidth,angle=0,trim = 0mm 0mm 0mm 0mm, clip]{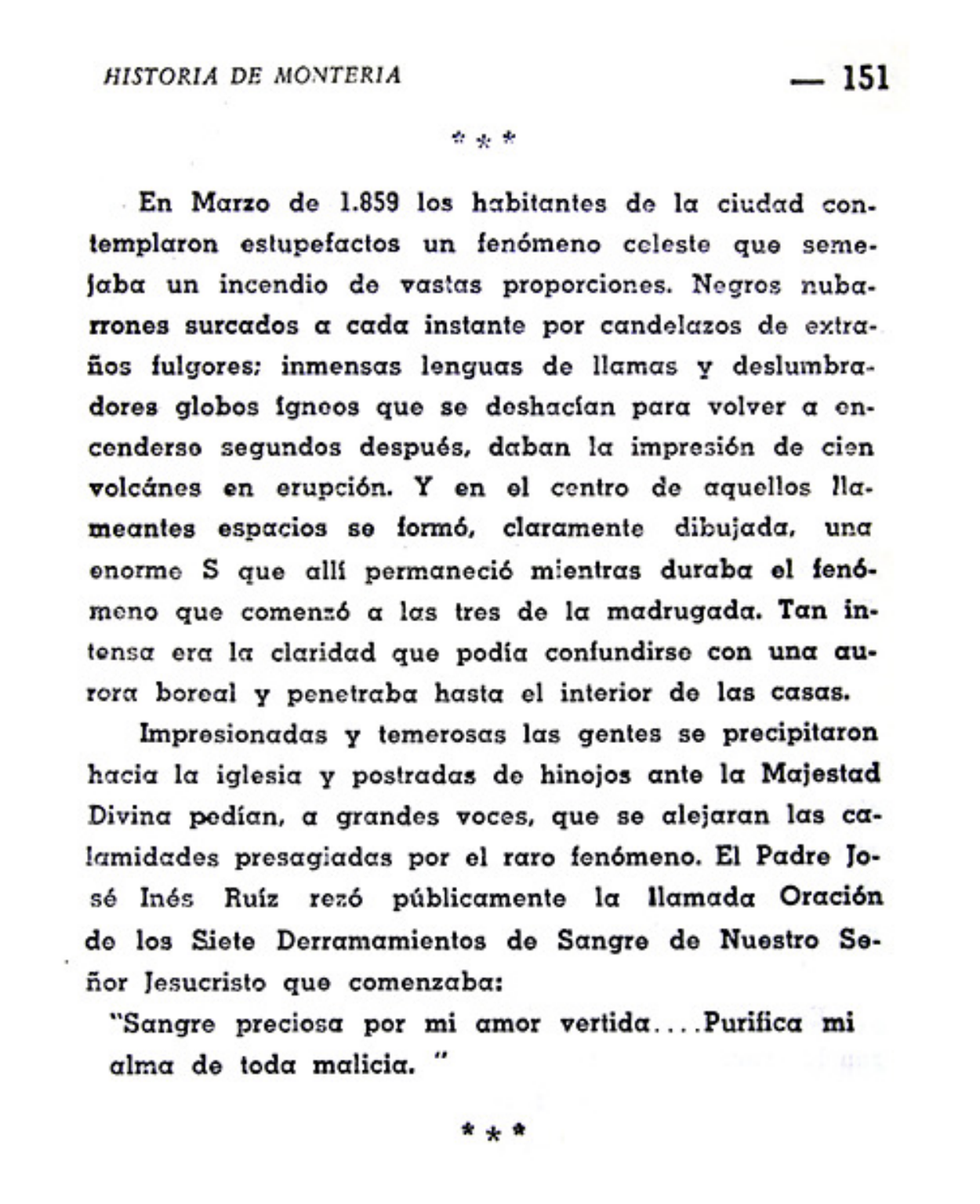}
       \caption{Original page from \cite{Exbrayat} describing (in spanish) a phenomenon likely to be attributed to the Carrington storm/event, based on observations from Monter\'ia, Colombia. The last sentence recites:  "Precious blood for my love spilled ...  Purify my soul from all malice". See the text for the complete english translation.}
     \label{figure3}
  \end{center}
\end{figure*}

A review of historical documents in an expedition carried out at the northern part of Colombia, allowed the identification of a description of the phenomenon in Monter\'ia (department of C\'ordoba), located at 8\grado 45' N. The report of this exceptional situation (shown in Fig.~\ref{figure3}) was most likely done, by the priest Jos\'e In\'es Ruiz, vicar of the town at that moment, as referenced by \cite{Exbrayat}: \\

  \emph{In March of 1859, the people of the town disquietingly envisaged a phenomenon of a vast proportion that resembled a great fire. Pitch-black storm-clouds furrowed by blazes of strange resplendence; immense flaming tongues and blinding igneous globules that disappeared only to light back seconds after, giving the impression of a hundred erupting volcanoes. And, in the midst of those flaming spaces, a humongous S was clearly drawn in the sky, and it remained for as long the phenomenon lasted, beginning at three in the morning. The clarity was so intense, that it could have been confused for the northern lights, penetrating inside the homesteads.
Startled and timorous, the people ran to the church and prostrated kneeling before Divine Majesty praying out loud for the banishing of the calamity that the rare phenomenon foreshadowed. Father Jos\'e In\'es Ruiz appealed to the prayer of the seven spills of the blood of our Lord Jesus Christ ...}
  
Taking into account the previous descriptions, in this work we develop a comparison of the claims made by the above report and the commonplace phenomena of aurorae. Additionally, we aim at seeking for evidence that could validate the narrative, letting this description to be the lowest-latitude observation of the Carrington's event aurorae.

\section{On the different aspects of aurorae}

\subsection{Geomagnetic storms}

In general, various phenomena produced by solar activity are likely to cause disturbances of the Earth's magnetosphere, known as geomagnetic storms. The swedish astronomer  Olof Petrus Hiorter observed aurora phenomenon and noted simultaneously a great movement of a magnetic needle on March 1, 1741 \citep{Chapman2}. Later, on 21 December 1806, Alexander von Humboldt detected malfunctioning of a magnetic compass during an auroral event in Berlin, giving also a direct indication of a connection between geomagnetic storms and aurorae. Nowadays, it is already well known that geomagnetic storms are mainly caused by solar phenomena traveling through interplanetary space and reaching Earth's magnetosphere.

In summary, two main solar triggers of geomagnetic storms can be listed \citep{Beck}:

The first is related to solar flares, eruptive filaments, coronal mass ejections (CMEs), and other kinds of transient explosive activity that depend on the sunspot cycle. These events release solar plasma, that, when having magnetic fields directed opposite to the Earth's fields and striking the magnetosphere, results in both dayside and subsequent night side reconnection that can lead to acceleration of the particles that produce the aurora. Transitory explosive solar phenomena are fairly strong and could produce, in general, a very intense geomagnetic activity and auroral displays.

In second place, high-speed streams from coronal holes in the solar atmosphere (areas in which the solar magnetic field is opened and from where high-energy charged particles escape as jets traveling outwards from the solar surface) can produce auroral displays, but less intense than the ones generated by the first phenomena, and are not able to reach mid and/or low latitudes.

The triggers have actually different effects on geomagnetic storms. Flares may produce the 'crochet effect' mentioned before; filaments and CMEs (different counterparts of the same event, as magnetic flux rope ejections) may be the trigger of geomagnetic storms, but also coronal holes and their high-speed streams. These different solar features may form some distinctive features in ground magnetograms (crochet, CME pass, high speed streams usually form alfv\'enic disturbances). The frequency of these events is linked to the solar cycle, being slightly more frequent in the decline phase of the cycle. 

Currently, geomagnetic storms are recognised as natural hazards and their impact on Earth is a matter of study to asure normal activities in modern technology-dependent societies \citep{Cid}.

\subsection{The polar aurorae}

Aurorae, also known as Northern or Southern Lights, depending on their location at northern or southern latitudes respectively, are the visible manifestation of atmospheric phenomena in the night sky in the form of resplendent arcs, moving luminous rays, and other colored structures, mainly in red and green. Aurora comes from the Latin aurora, meaning dawn, because of the resemblance of the lights of the phenomenon and the sun's early daylight\footnote{In spanish, the same word is used for aurora and dawn.}. Polar aurorae are produced by charged particles  reaching Earth from outer space (mainly from the Sun) and interacting with the atmosphere of our planet, as commented in the previous section. They can collide with the atoms and molecules in the upper atmosphere, causing ionization and excitation of atmospheric constituents with the subsequent light emission \citep{Lilensten}. Part of the energy is released as visible light \citep{Beck}.

Located mainly in altitudes of 100 km, aurorae form constantly changing irregular ovals, centered in the magnetic poles of the Earth. Aurorae are mainly produced by plasma clouds (CMEs) or high-speed streams as previously commented. Low-latitude aurorae such as the Carrington event are primarily due to very energetic CMEs. The bombing of high-energy particles is organized in beams and focused by the magnetic fields in the upper atmosphere, creating spirals in the lines of the terrestrial magnetic field directed towards the poles. 

In 1722, George Graham discovered short-term variations in the magnetic field, some of them were regular and occurred within the course of day, but others were anomalous perturbations. In 1740, Anders Celsius understood that these variations happened simultaneously in two distinct locations, hypothesizing that the terrestrial magnetic field seemed to be altered in a general scale, rather than locally. Then, in the XIX Century, the German physicist K.F. Gauss discovered that the terrestrial magnetic field originated in the interior of the planet, instead of the atmosphere. Current theories postulate that the magnetic field is generated by cells of convection in the Earth's liquid nucleus, constituting a dynamo. While these breakthroughs occurred, research on aurorae continued, improving the understandings of their geographical distribution. It was evident that there was an increase in the number of sightings towards the poles (in the southern British Islands, on average, there were six observations each year, while farther in the north, there were 60). The auroral region is an area bounded by two ovals (north and south) where the displays are observed with more frequency, firstly demonstrated by \cite{Feldstein}.

During the most intense geomagnetic storms, the auroral oval \citep{Phillips} expands towards the equator, turning visible in tropical latitudes \citep{Gonzalez}. The sightings of polar aurorae in low latitudes are rare and only four times in the last 160 years they have been studied/reported : 1859, 1872, 1909 and 1921 \citep{Silverman}.

Due to the geomagnetic secular variation, the Earth's magnetic field changes (in strength and direction).  The dipolar component is dominant and determines the direction of the magnetic poles, which migrate on time scales of a year or more. The magnitude associated to the magnetic field varies between 0.6 Gauss near to the poles, and 0.3 Gauss near to the equator. Research on magnetized rocks reveal that the complete magnetic field has changed its direction over twice every million years in the last 165 million years \citep{Daintith}.

\begin{figure*}
  \begin{center}
    \leavevmode
      \includegraphics[width=0.527\linewidth,angle=0,trim = 0mm 0mm 0mm 0mm, clip]{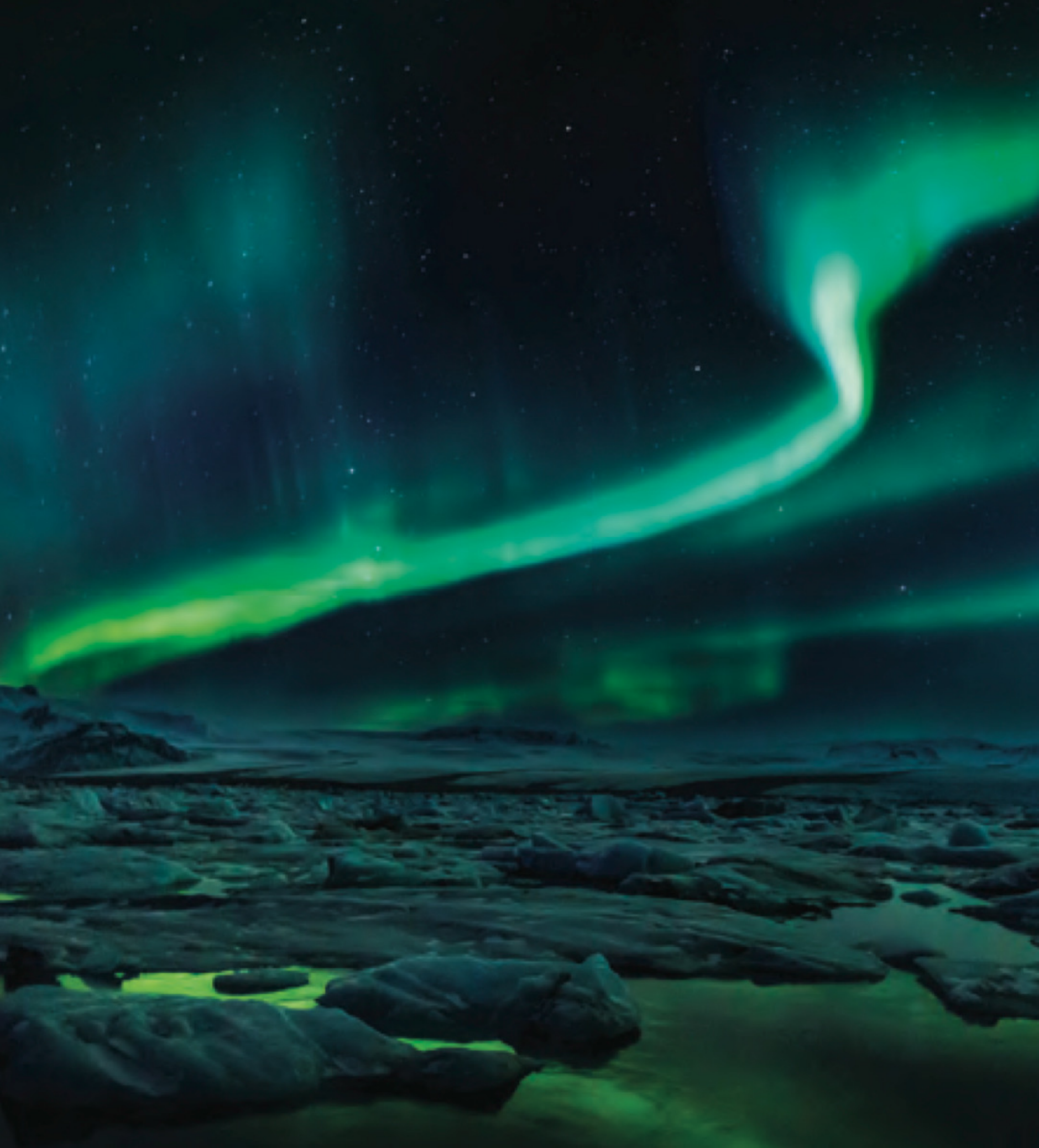}~~\includegraphics[width=0.44\linewidth,angle=0,trim = 0mm 0mm 0mm 0mm, clip]{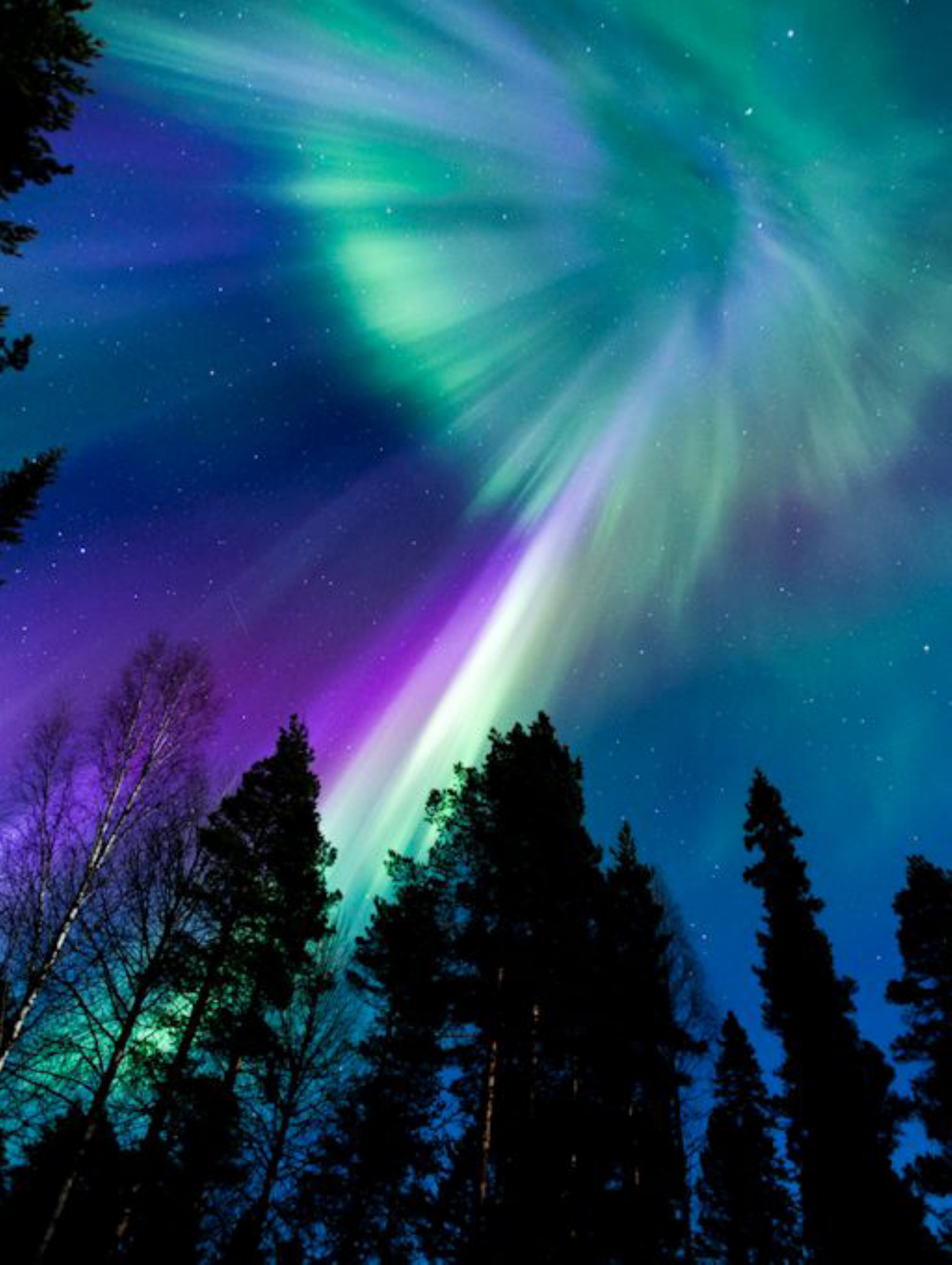}
       \caption{Aurorae observations with different shapes. \emph{Left}: Homogeneous Band Aurora, resembling the one observed in Monter\'ia in 1859.  \emph{Right}: Coronet Aurora. Taken from Markku Inkil  (\url{https://www.facebook.com/pages/Revontulet/197376608535}).}.
     \label{figure4}
  \end{center}
\end{figure*}

The spectacular beauty of an aurora defies its description. Simple words fail to completely transmit the sensation of amusement and majesty.  Classifying aurorae according to their appearance, without explanation of their cause, there are three main types of shape:  structures without rays (HA: homogeneous arc, HB: homogeneous band, PA: pulsating arc, DS: Diffuse Luminous Surface, PS: pulsating surface, G: gleams), structures with rays (R: rays, RA: rayed arcs, RB: rayed bands, D: drapers, C:coronets, F: radiant coronets or fans) and shining forms, as described in \citet{Petrie, Eather}. Figure~\ref{figure4} displays quite common aurorae shapes.

Since aurorae are caused by gaseous discharges of light, emissions occur in discrete wavelengths. From upper to lower atmospheric heights, and due to the different components and gas densities, the emissions exhibit different characteristic colors. See \cite{Beck} and references therein for a detailed description of corresponding emissions.

Regarding the human perception of aurora colors, on average, because of the intensity of the aurora, green and red can mix along with the light violet emitted by simple ionized nitrogen (3914 $\AA$) in order to produce greyish yellow tones. Generally, the color of aurorae, organized on their rareness: green, white, red, blue and yellow. The lines of oxygen are one of the highest in intensity, but the eye is less sensitive to this wavelengths, making the emissions look tenuous. The color white is often associated to the scotopic vision of the aurorae due to the low levels of illumination when the aurora is faint and the cones provide of a light that does not stimulate the eye's \citep{Beck}.

\subsection{Black Aurorae}

\begin{figure*}
  \begin{center}
    \leavevmode
      \includegraphics[width=0.9\linewidth,angle=0,trim = 0mm 0mm 0mm 0mm, clip]{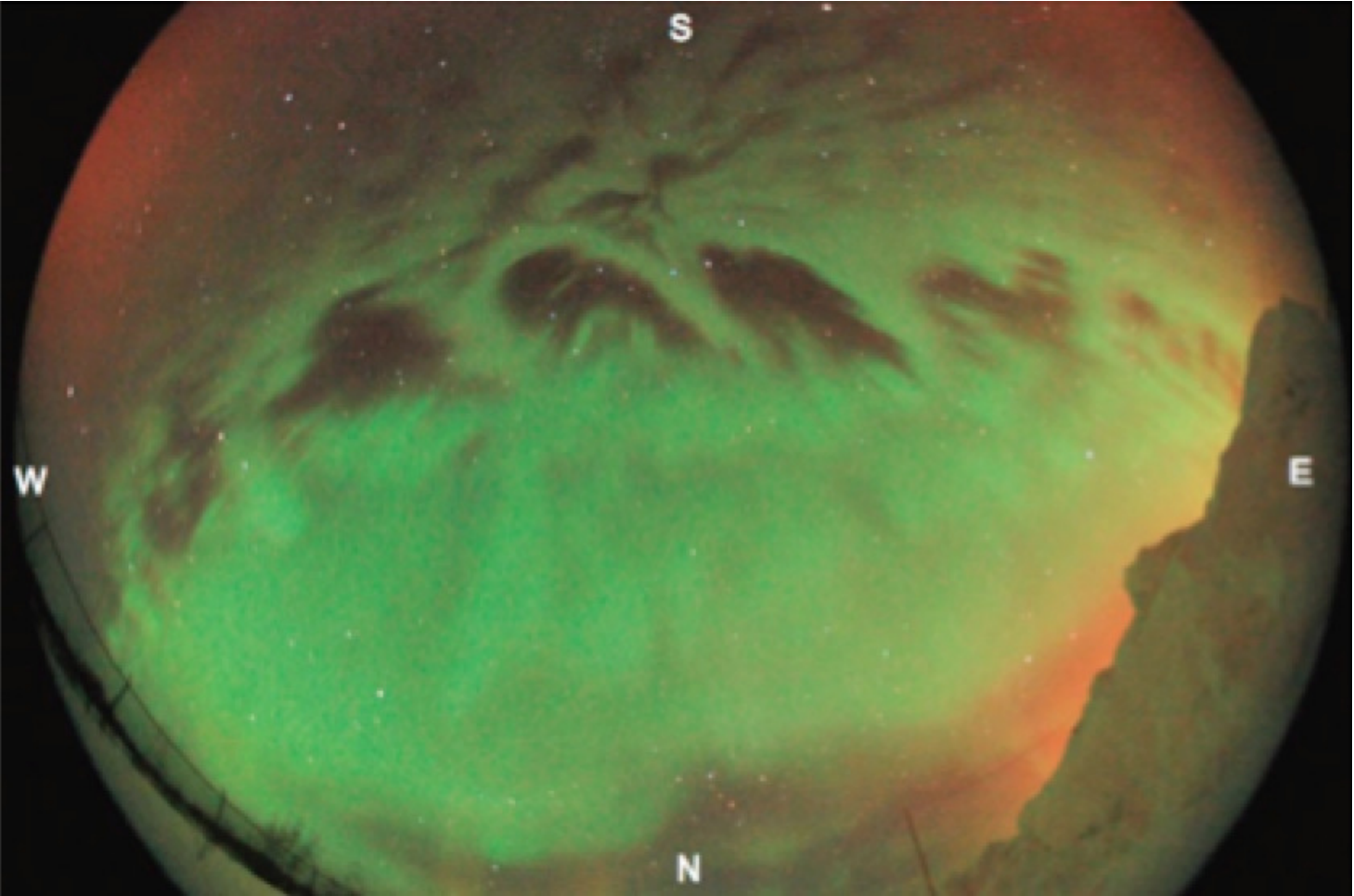}
       \caption{The Black Aurora. Taken from \cite{Sagakuchi}.}
     \label{figure5}
  \end{center}
\end{figure*}

\begin{figure*}
  \begin{center}
    \leavevmode
      \includegraphics[width=1.0\linewidth,angle=0,trim = 0mm 0mm 0mm 0mm, clip]{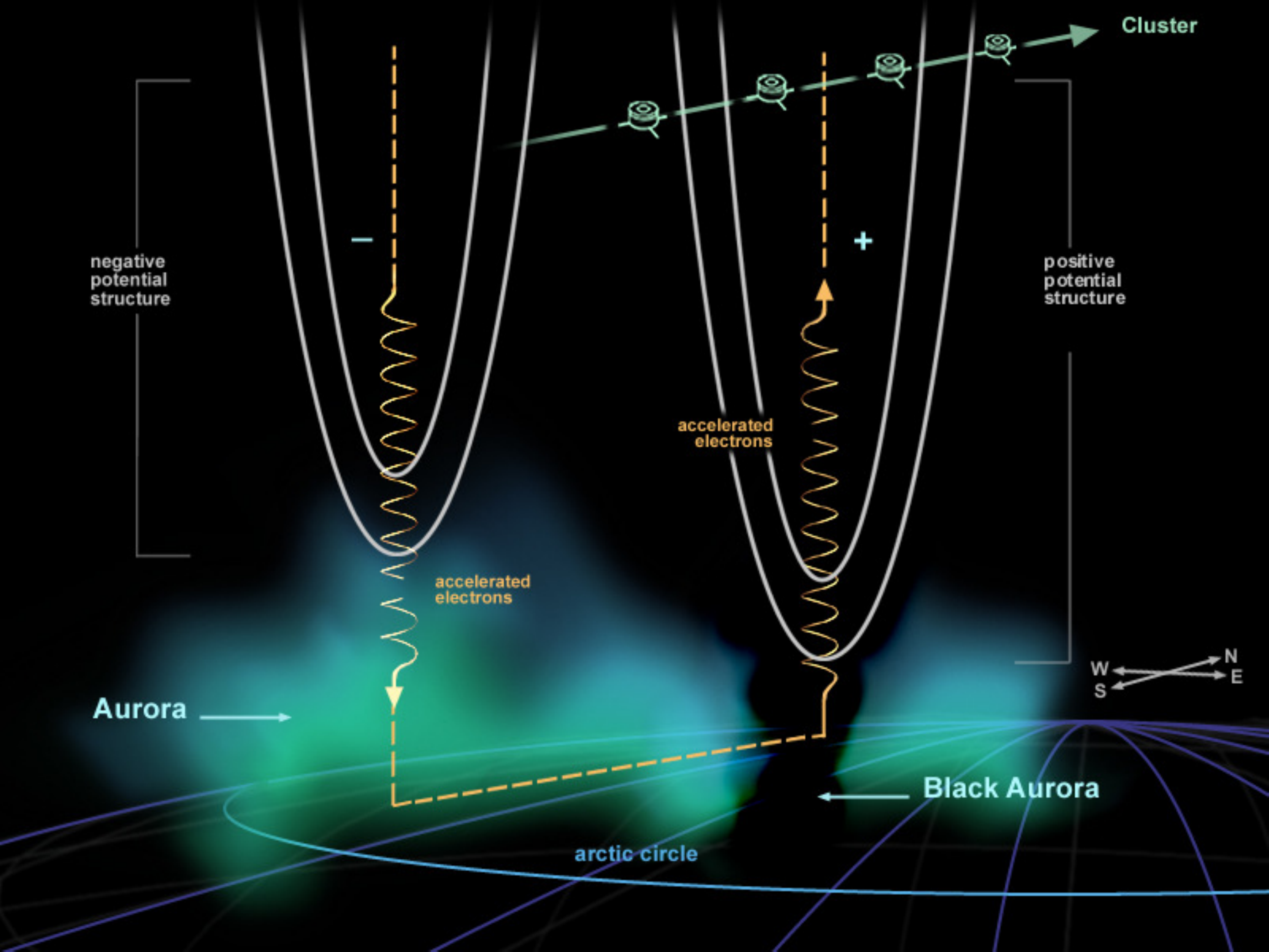}
       \caption{The black aurora constitutes a region where the descending particle currents close the aurora's current circuit, then, electrons flow outwards from the ionosphere. Taken from the images of the ESA's Cluster Mission (\url{http://www.esa.int/Our\_Activities/Space\_Science/Cluster/Cluster\_quartet\_probes\_the\_secrets\_of\_the\_black\_aurora)}.}
     \label{figure6}
  \end{center}
\end{figure*}

When the sky is all covered in aurorae, like a veil, thin regions and "patches" without luminosity appear (Fig.~\ref{figure5}). Features move and give the visual impression of a black aurora \citep{Eather}. These darkened regions own a dynamic characteristic similar to those of arcs in a small aurora, a different scale, yet, the rotation sense of the black loops is opposite to the aurora's. Recent theories explain the phenomenon as a region in which the descending particle currents close the aurora's current circuit, henceforth, electrons flow outwards from the ionosphere (Fig.~\ref{figure6}),  \citep{Blixt, Archer}. 

It has been studied that in the black aurorae (black arcs and black patches) are not associated with a field-aligned potential (e.g. divergent potential structure) but are likely to be caused by the suppression of pitch angle scattering \citep{Sakanoi}. An overlap of low-energy (2 to 5 keV) electrons traveling downwards and high-energy (several keV) plasma sheets is commonly detected in the black regions.  See \cite{Sakanoi}, and references therein, for a more detailed discussion on the generation process of the black aurora.

Based on different observation, some authors, e.g. \cite{Trondsen, Kimball}, have associated black aurorae with pulsating aurorae. The pulsating one has not received the same attention as other types of aurora though recent results evidence that it is much more widespread than previously thought, i.e., more important in terms of magnetosphere-ionosphere coupling, as recently reviewed by \cite{Lessard}. The description by \cite{Exbrayat} of the historical report in Colombia, coincides with the representation of both, the black and the pulsating aurorae, and evidenced through the following lines: \emph{Black clouds plowed from time to time by flashes of strange glint.} 

\subsection{Low-latitude aurorae}
To date, the aurorae seen in September, 1859, is recognized as arguably the most intense in the last 160 years, being reported in Cuba and Panama \citep{Green2006}; February 4, 1872, another great aurora was registered in Bombay, India (19\grado N), and one more in May 15, 1921, in Samoa (14\grado S) \citep{Eather}. Even if the latitudes mentioned before are geographic, aurorae are controlled by the Earth's magnetic field. The geomagnetic equator is 17\grado (maximum) misaligned in South America. Despite geographic latitudes in Colombia (8$^\circ$ 45' N) and southern Japan (30$^\circ$ 20' N) are considerably different, their corresponding geomagnetic latitudes are quite similar, i.e., Monter\'ia, Colombia (18$^\circ$ 17') and southern Japan (21$^\circ$ 12' N), due to the displacement of the geomagnetic equator towards South America. Consequently, there is a higher chance for aurorae to be sighted in this region than in the southern part of Japan, and the aurorae will extend following the geomagnetic latitude parallels (Fig.~\ref{figure7}). It should be noted that Japan has a large latitudinal extension and that, although these auroras were observed in northern Japan, it seems that the sky was very overcast at that time in the south \citep{Green2005a}.

This particular situation validates the report made by \cite{Exbrayat}. Monter\'ia is 8.75\grado N, and its geomagnetic latitude is 18.86\grado N. The neighboring country of Panama, has a geographic latitude of 9\grado N, and a geomagnetic latitude of 20.12\grado N  (in 1900), impressively close to Monter\'ia's, below the 18\grado N limit proposed by \citep{Green2006} for the great aurora of 1859. Table~\ref{table} shows the variation of geomagnetic latitude in over a century, and the geographic latitude for both cities.

\begin{table}
\centering
\begin{tabular}{|lccc|}\hline
 Location & Geographic lat. & Magnetic lat. & Magnetic lat.\\
 &  & (year 1900) & (year 2015)\\\hline
Monter\'ia & 8.75$^\circ$ & 19.99$^\circ$& 18.37$^\circ$ \\
Panama City & 9.00$^\circ$ & 20.12$^\circ$ & 18.56$^\circ$\\\hline
    \end{tabular}
  \caption{Geographic and geomagnetic latitude (according to the International Geomagnetic Reference Field, IGRF-11 model for coordinate transformation) for Panama City (Panama) and Monter\'ia (Colombia).}
    \label{table}
\end{table}

\section{Analysis and Discussion}

In the following section the narrative described by \cite{Exbrayat} will be analyzed, looking for evidence that allows the validation of this phenomenon as an aurorae sighting. 
The first lines of the report make allusion to a particular date: 
\emph{In March of 1859, the people of the town disquietingly envisaged a phenomenon of a vast proportion} \citep{Exbrayat}. This mentioned date is March of 1859, though the great aurorae occurred in the beginning of September. In order to verify this data, a visit to the city of Monter\'ia was made in May, 2014. with the purpose of looking in the different libraries of the town, especially the Diocese's archive, as the report references father Jos\'e In\'es Ruiz as a witness. During the expedition to the northern regions in Colombia and in a particular visit to Monter\'ia's Cathedral and its Diocese, annotation books or booklets written by the father were searched, although, none was found, henceforth, the reference on which \cite{Exbrayat} supported his report is unknown.
However, after carefully reviewing the birth and marriage registry books, it was corroborated that these had been written by father Ruiz himself. In the first book that was checked, some drawings were found in the contour of the title given to the registers of September 1859, clearly depicting diverse forms similar to spirals and S-shapes of aurorae. Homologous titles in previous months (for a register that represented almost 20 years) had no drawing at all, suggesting that the mistaken date in \cite{Exbrayat} was probably due to an error in the historian original reading of the document because of its antiquity. Figure~\ref{figure8} shows particular pages of the birth and baptism registry book from Monter\'ia's Cathedral (original and enhanced images) with the corresponding pages for August and September. It can be noted a plain text for August (as it is for previous months), in contrast with the features that garnish September, highlighted with arrows in the figure. Figure~\ref{figure9} (left) shows a photography of the spiral shape of an aurora that is further extracted in the panel a) on the right. Panel b) in the same figure shows the enhanced shape drawn by father Ruiz in the birth and baptism registry book of September 2, 1859 (see the large arrow in Fig.~\ref{figure8}) that resembles the shape in panel b).

Furthermore, and supporting our interpretation, in an examination looking for reports on auroral displays, there are not great storms listed for March 1859 \citep{Jones, Nevanlinna}.

The auroral activity of 1859 began in the night from the 28th to the 29th of August although, it was seen as far as northern Cuba \citep{Green2006}. Henceforth, it is very likely that the habitants of Monter\'ia observed the aurora, during the dawn on September 2, that particularly matches with the first baptism celebrated in the first church of Monter\'ia, cathedral nowadays.

The moon was in the rising phase in Monter\'ia by the beginning of September in 1859, with moonset at 21:40 (local time), thus allowing good sky conditions in terms of darkness to have a good visibility of luminosity enhancements caused by auroral activity.

\begin{figure*}
  \begin{center}
    \leavevmode
      \includegraphics[width=1.0\linewidth,angle=0,trim = 0mm 0mm 0mm 0mm, clip]{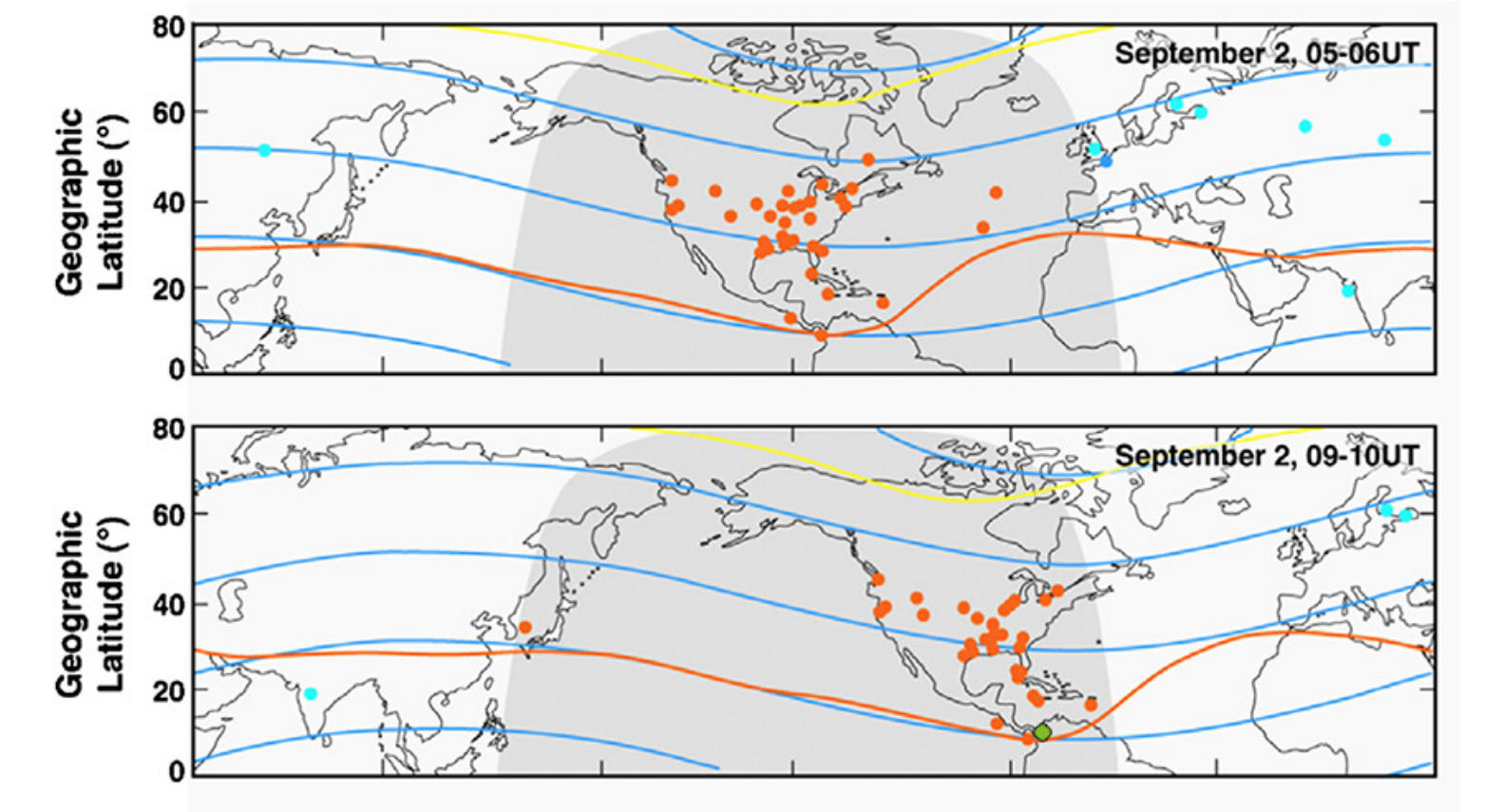}
       \caption{Locations where the great auroral activity on September 2, 1859, was sighted and reported (orange dots), starting 3:00 UT) as sketched by \cite{Green2005a}. Magnetometer stations are also included  (blue dots). These authors drew the  geomagnetic dipole latitude with blue lines with the yellow and orange lines representing minimum and maximum extent of the auroral oval from Holzworth-Meng model, respectively. In this work we superimposed the green dot to highlight the location of Monter\'ia (Colombia). The position of the orange dot that corresponds to the Panama report in \cite{Green2005a} would actually be slightly higher up than the position of our new detection in Monter\'ia, Colombia.}
     \label{figure7}
  \end{center}
\end{figure*}

\begin{figure*}
  \begin{center}
    \leavevmode
      \includegraphics[width=1.0\linewidth,angle=0,trim = 0mm 0mm 0mm 0mm, clip]{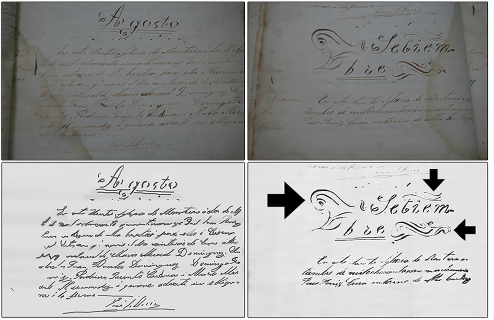}
           \caption{Extracts from birth and baptism registry book from Monter\'ia's Cathedral on August (\emph{left)} and September (\emph{right}) in 1859. Original documents (\emph{upper panels}) and enhanced copies (\emph{lower panels}) are shown. Arrows point out the various markings interpreted as sketches of the aurora. \emph{August: In this holy church of Monter\'ia, at two of August 1859, I Jos\'e In\'es Ruiz, interim priest of it, baptized, put oil and chrism to Jos\'e Esteban, who was born on the last January 26, the illegitimate son of Mar\'ia Merced Dom\'inguez . His grandparents are Juan Bacilio  Dom\'inguez and Dominga Fern\'andez. Godmothers are Jacinta C\'ardenas and  Mar\'ia Merced Hern\'andez and Jim\'enez. I remarked their duties and signed,  Jos\'e I. Ruiz. September: In this holy church of Monter\'ia, at two of September 1859, I Jos\'e In\'es Ruiz, interim priest of it, baptized Manuela de Febres.}}
     \label{figure8}
  \end{center}
\end{figure*}

\begin{figure*}
  \begin{center}
    \leavevmode
      \includegraphics[width=0.8\linewidth,angle=0,trim = 0mm 0mm 0mm 0mm, clip]{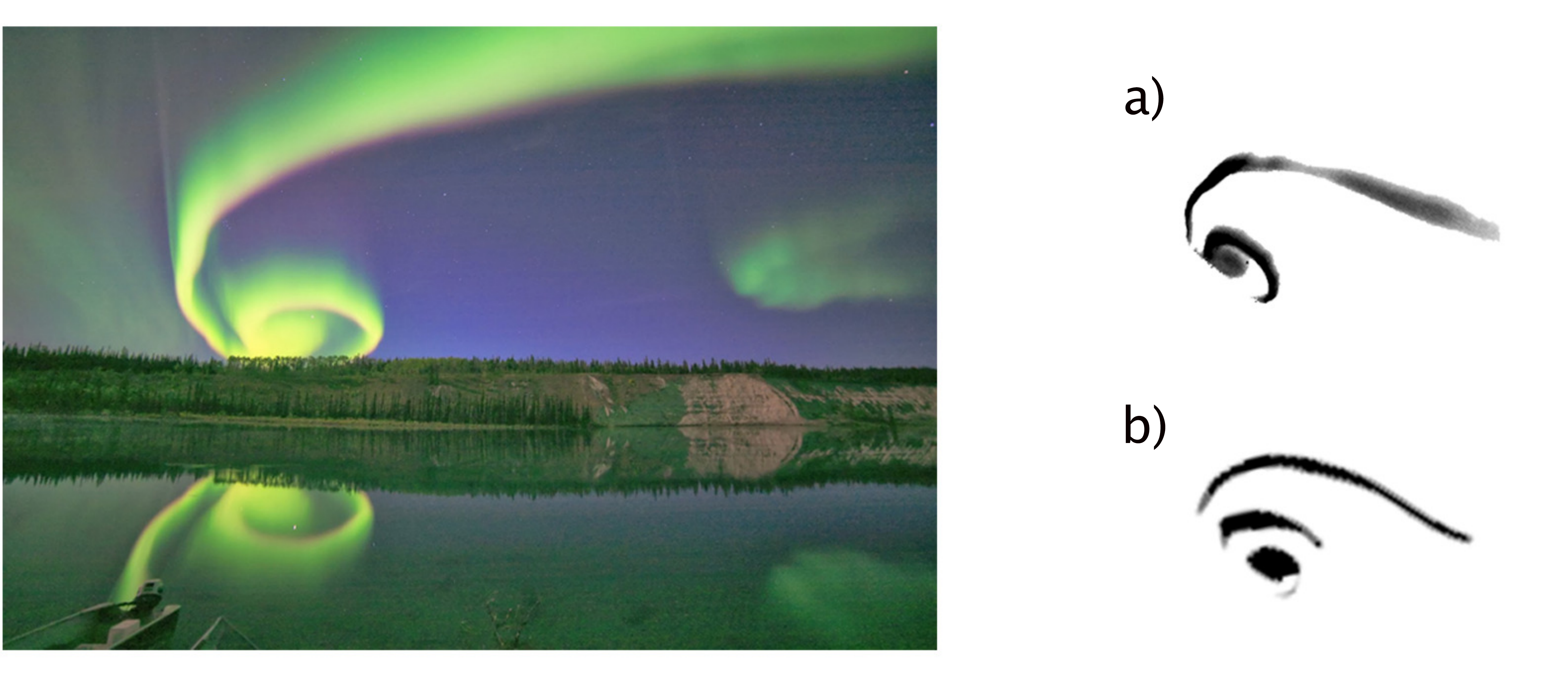}
       \caption{\emph{Left}: Photography of an aurora borealis (September, 2012; Yukon Territory) with a curly shape taken by David Cartier (\url{https://www.flickr.com/photos/dcartiersr/7927971874/}). \emph{Right panels}: a) Spiral shape of the aurora extracted from the picture on the left. b) Enhanced image of the drawings of father Ruiz in the birth and baptism registry book of September 2, 1859 (see large arrow in Fig.~\ref{figure8}).}
     \label{figure9}
  \end{center}
\end{figure*}

Regarding the fact that it was a \emph{phenomenon of a vast proportion that resembled a great fire}, it is important to recall that this description is common in aurorae reports of different eras. Seneca once narrated that a group of soldiers, dispatched to the Roman colonies in Ostia, envisaged an aurora which was mistakenly interpreted as a fire \citep{Eather}. 
An office of the Associated Press of London, described regarding the sighting of a bright aurora in January 25th 1938:
\emph{The reddy glow led many to think. The Windsor fire department was called out in the belief that Windsor Castle was afire. In Austria and Switzerland, firemen turnout to chase nonexistent fires.} \citep{Eather}.

Furthermore,it has been recalled that in the regions where the aurorae of 1859 was seen, the local people believed their cities were on fire. During the aurora observed the night from the 14th to the 15th of May, 1921, many accounts of fires in Arizona, Mexico and Jamaica were communicated. Also, a signature characteristic of low-latitude aurorae is a reddish coloration \citep{Chapman,Green2005a}.

The following sentence has an intriguing description: \emph{Pitch-black storm-clouds furrowed by blazes of strange resplendence $\cdot$}
It is reasonable to understand that the first phenomenon mentioned in the previous sentence refers to the black aurora, seen also in Boston in the same date \citep{Green2006}:  \emph{About 10 [PM]\footnote{In local time. Standard time was not enacted into US law until 1918.} a tremulous flashing up from the east was observed - soon after a bank-like arc of a circle was seen in the North, below which, the appearance was very somber, resembling a very dark cloud. From this arc soon shot up columns of light toward the zenith. This was immediately succeeded by the most lively and brilliant succession of flashes ...} [Boston Transcript, Saturday, September 2,1859]. The black aurora was likely to be observed in Monter\'ia, according to the descriptions of father Ruiz, but about 5 hours after the Boston report. It should be noted that both cities lay almost along the same meridian: Boston 71.05$^\circ$W, Monter\'ia 75.88$^\circ$W.

\begin{figure*}
  \begin{center}
    \leavevmode
      \includegraphics[width=1.0\linewidth,angle=0,trim = 0mm 0mm 0mm 0mm, clip]{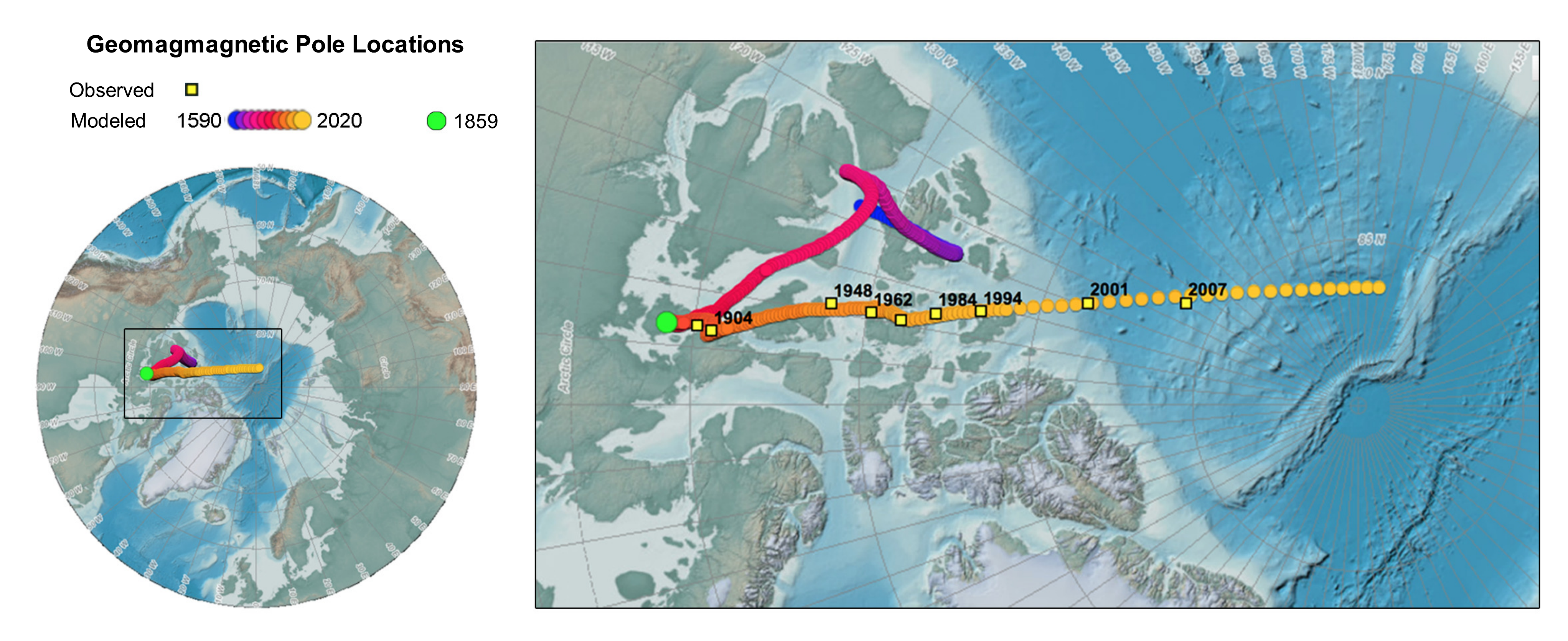}
       \caption{Historical magnetic declination (Arctic). The geomagnetic north pole locations are marked according to the information and color-code shown in the figure (numbers represent years). Composite image based on the information from observations and models (\url{http://maps.ngdc.noaa.gov}). Note that the southernmost location of the magnetic north pole (green dot) in its secular migration in almost five centuries, occurred in 1859 and is located at  (69.174$^\circ$N, 96.757$^\circ$W).}
     \label{figure10}
  \end{center}
\end{figure*}

An important fact we want to stress in this work, and that could be one of the factors explaining the visibility of aurora displays at low latitudes in 1859, is related to the geomagnetic secular variation. Figure~\ref{figure10} shows the secular variation of the north magnetic pole in almost five centuries. It can be evidenced that, among all locations of the geomagnetic north pole, its latitude in 1859 is the lowest (69.174$^\circ$N, 96.757$^\circ$W). Other factors are proposed in the literature, e.g related to plasma injection into the nightside magnetosphere, superposition of magnetopause currents and partial ring current,  as further explained in \cite{Cid} and references therein.

Black aurorae were previously defined as structures surrounded by the luminosity of the aurora, in which there is an absence of emission \citep{Sagakuchi}. There seems to be two different types of black aurorae. The key difference between both types lies in the fact that the first type, that encompasses black spots, arcs and rings, does not show any kind of underlying cut mechanism, while the second type, consisting in lines of black vortices and arcs with curling ends, exhibits a clear evidence of shear \citep{Archer}. 

Often their shape is elliptic and it has been observed that it can develop into segments and rings, and backwards. Blackened spots have been seen to move in groups eastwards, and rarely westwards. \cite{Kimball} calculated that their velocity was approximately 1.5 km s$^{-1}$. The "\emph{immense flaming tongues} probably correspond to the greenish yellow lights that are generated in heights of 150-90 kilometers, due to the interaction of the denser layers of oxygen with the particle currents, emitting in 5577 $\AA$. And the \emph{blinking igneous globules that disappeared only to light back seconds after} can be related to the pulsating surfaces of light patches, which appear and disappear in the same spot periodically in a few seconds \citep{Chapman}.

\emph{And, in the midst of those flaming spaces, a humongous S was clearly drawn in the sky, and it remained for as long the phenomenon lasted, beginning at three in the morning.} This is perhaps one of the most interesting annotations matching with a description of an homogeneous band aurora, and the depiction of the drawings in the registry book.
The crew of the Saranac in Panama (Fig.~\ref{figure7}) perceived the aurora at four in the morning, on August 28 \citep{Green2006}, suggesting that this phenomenon could have also been sighted in Colombia that very same night. 

\cite{Kimball} reconstructed the auroral displays observed in Havana (Cuba), at different times, based on the report by \cite{Loomis}.  According to \cite{Kimball2}, at 8:00 UT (september 2nd) , three different phenomena were observed: homogeneous arc, red aurora and glow. The description of father Ruiz in Monter\'ia,  resembles the same characteristics reported in Cuba, and with a good agreement in terms of timing. The blinking igneous globules described by father Ruiz are likely to be referred to some spacial cases of pulsating aurora with bursts of luminous \citep{Eather}.

Yet it is more likely that given phenomenon corresponds to the dawn of September 2nd, because of the pictures found in the birth and baptism registry books.
The following description, recounted by Exbrayat, implies that the one that envisaged the aurora and reported back to him, at least knew its characteristics because of the explicit reference to the northern lights: \emph{The clarity was so intense, that it could have been confused for the northern lights, penetrating inside the homesteads.} The luminous intensity of the phenomenon is often misinterpreted by a first observer as the lights of dawn, as evidenced in some reports of this event, e.g \emph{"... the light observed during the night was so intense that it woke up many people in different places, such as it occurred in Cuba."} and \emph{"On 2d September, an Aurora Borealis was seen at Guadeloupe\footnote{West Indies (lat. 16$^\circ$ 12').} ... its ruddy light was noticeable in the interior of the houses. The aurora stained its maximum of brightness at 3 A.M. (local time)"} cited by \cite{Silliman}, and in other reports of auroral activity \citep{Green2006}.

\section{Final Remarks}

The solar active region 520 constitutes the origin of the great auroral activity of 1859, its size was of 2300 millionths of a solar hemisphere, affecting other aspects of space weather like: sudden ionosphere alteration, solar energetic particles and solar wind. In particular, the probable CME ejected from that active region is likely to be responsible for the experienced consequences on Earth. The Carrington Event affected almost half of the telegraphic stations in the United States. The first network of telegraphs in Colombia, connecting Bogot\'a with Nare (Antioquia), 180 km, was installed in 1865 \citep{Rodriguez}, six years after the Carrington Event. Research conducted by the National Academy of Sciences of the United States predicts that the damage caused by a similar phenomenon today would be comparable to that of twenty hurricanes Katrina \citep{NatResCoun}.

The description in \cite{Exbrayat} corresponds to the sighting of the aurora borealis of August and September of 1859 due to the following:

\begin{enumerate}
\item Although the date given by \cite{Exbrayat} is not accurate, the drawings found in the birth and baptism registry books of September 1859 are clearly suggesting the observation of the great auroral activity of that month. The apparent data mismatch is likely to have occurred because of the difficulties implicit in the handling of antique documents. The choice of September over March is supported by the absence of reports of any large storms in March 1859.
\item The narrative of the event occurred in Monter\'ia describes a variety of phenomena associated with aurorae: intense illumination of the atmosphere, the appearance of a bright S-shaped homogeneous band, and the description of black aurorae.
\item The geomagnetic latitude of Monter\'ia (19.99\grado N) by the end of the 19th century was particularly close to that of Ciudad de Panam\'a (20.12\grado N), where there is known evidence of the aurorae of 1859. The description given by \cite{Exbrayat} would represent the lowest latitude report of aurorae of the Carrington Event, being also one that is impressively complete and detailed.
\item The time in which the Monter\'ia's aurora was observed (3:00 A.M., local time) matches the time to establish that it was the same auroral event extensively reported in higher latitudes, e.g. North America.
\item Other reports, similar to that of Monter\'ia, are likely to exist in the North of South America and Southern Central America, according to \cite{Green2006} who suggest that the aurora of September 1st and 2nd, 1859, extended as far as 18\grado geomagnetic North. The very low latitude of the magnetic north pole in 1859 caused by the geomagnetic secular variation is very likely to be responsible of the effects on Earth occurred during the Carrington Event, in particular the low-latitude aurora seen in 1859. In a future work we will discuss on the relation between the location of the magnetic poles and major auroral events.

\item The actual relationship between black aurora and pulsating aurora is still under debate \citep{Sakanoi}. The description found in the report from Monter\'ia supports the simultaneous presence of both.
\end{enumerate}

As a final anecdote and matter of curiosity, the lyrics of the Colombian National Anthem include in one of the stanzas the following sentence: \emph{Sublime freedom
spills the dawns of its invincible light}. Noting that dawn and aurora are the same word in spanish, we could interpret the phrase as describing auroral events, which, so far, has never been the interpretation of it. However, the preceding word "spills" does not indicate the idea of "rising", as it should be the common and expected description of the dawn, but for "falling", as certainly aurorae activity is generally reported. In a  deeper exploration of the plausible connection with truly auroral descriptions, it should be pointed out that the lyrics were written by Rafael N\'u\~nez, writer, politician and president of Colombia in four periods during the 19th century. N\'u\~nez was also elected Governor of the state of Panama (one state of the United States of Colombia in 1859) in 1858. Based on historical evidence found in \cite{Lievano}, it can be established that in September of 1859 N\'u\~nez was in Panama, and we therefore argue that it could have been possible for him to observe or, at least, had news of the reports, of the auroral displays that are known to have occurred in Panama. Further exploration in the historical documentation and works of N\'u\~nez brought us to find three poems \citep{Nunez} in which he referred to aurorae, giving us some signs to speculate that the lyrics in the Colombian National Anthem are truly giving mention to the observation of the Carrington Event. 

Acknowledgements:
The authors express their gratitude to Dr. Luis Enrique G\'omez Casabianca, father Ren\'e van Hissenhoven S. J., father Orlando L\'opez and the personnel of the Cathedral of Monter\'ia's Archive. We also thank the two anonymous referees for their important contributions and suggestions that helped to improve this paper.

\end{document}